\documentclass[aoas,nameyear,dvips]{arximspdf}
\usepackage{dcolumn}
\usepackage{graphicx}

\doi{10.1214/10-AOAS375}
\volume{5}
\issue{1}
\pubyear{2011}
\firstpage{427}
\lastpage{448}

\makeatletter

\newcolumntype{d}[1]{D{.}{.}{#1}}

\newcommand{\blam}{\bolds{\lambda}}
\newcommand{\bbeta}{{\bolds{\beta}}}
\newcommand{\bY}{\mathbf{Y}}
\newcommand{\bX}{\mathbf{X}}
\newcommand{\bc}{\mathbf{c}}
\newcommand{\A}{\mathcal{A}}
\newcommand{\bh}{\mathbf{h}}
\newcommand{\hbeta}{\widehat{\bbeta}}

\newtheorem{claim}{Claim}
\newtheorem{theorem}{Theorem}
\newproclaim{definition}{Definition}

\let\epsilon\varepsilon

\makeatother

\begin{document}
\begin{frontmatter}

\title{Improved variable selection with Forward-Lasso adaptive shrinkage\thanksref{TITL1}}

\thankstext{TITL1}{Supported in part by NSF Grants DMS-07-05312 and DMS-09-06784.}

\runtitle{Forward-LASSO with adaptive shrinkage}

\begin{aug}

\author[A]{\fnms{Peter} \snm{Radchenko}\ead[label=e1]{radchenk@usc.edu}}
and
\author[A]{\fnms{Gareth M.} \snm{James}\corref{}\ead[label=e2]{gareth@usc.edu}}

\runauthor{P. Radchenko and G. M. James}

\affiliation{University of Southern California}

\address[A]{Marshall School of Business\\
University of Southern California\\
Los Angeles, California 90089\\
USA\\
\printead{e1}\\
\phantom{E-mail: }\printead*{e2}} %adresu isvedimo komanda gale!
\end{aug}

% HISTORY:
\received{\smonth{12} \syear{2009}}
\revised{\smonth{6} \syear{2010}}

% ABSTRACT
%
\begin{abstract}
Recently, considerable interest has focused on variable selection
methods in regression situations where the number of predictors, $p$, is
large relative to the number of observations, $n$. Two commonly applied
variable selection approaches are the Lasso, which computes highly
shrunk regression coefficients, and Forward Selection, which uses no
shrinkage. We propose a new approach, ``Forward-Lasso Adaptive
SHrinkage'' (FLASH), which includes the Lasso and Forward Selection as
special cases, and can be used in both the linear regression and the
Generalized Linear Model domains. As with the Lasso and Forward
Selection, FLASH iteratively adds one variable to the model in a
hierarchical fashion but, unlike these methods, at each step adjusts the
level of shrinkage so as to optimize the selection of the next variable.
We first present FLASH in the linear regression setting and show that it
can be fitted using a variant of the computationally efficient LARS
algorithm.
%We then provide a theoretical
% analysis showing that some of the variable selection results that have
% recently been developed for the Lasso can be extended using FLASH.
Then, we extend FLASH to the GLM domain and demonstrate, through
numerous simulations and real world data sets, as well as some
theoretical analysis, that FLASH generally outperforms many competing
approaches.
\end{abstract}

% KEYWORDS
%

\end{frontmatter}

%s1 ###
\section{Introduction}
Consider the traditional linear regression model
%
%e1 ###
\begin{equation}
\label{linmodel}
Y_i = \beta_0+\sum_{j=1}^p
X_{ij} \beta_j + \epsilon_i,\qquad  i=1,\ldots, n,
\end{equation}
with $p$ predictors and $n$ observations. Recently attention has focused
on the scenario where $p$ is large relative to $n$. In this situation
there are many methods that outperform ordinary least squares (OLS)
[\citet{frank1}]. One common approach is to assume that the true number of
regression coefficients, that is, the number of nonzero $\beta_j$'s,
is small,
in which case estimation results can be improved by performing variable
selection. Many classical variable selection methods, such as Forward
Selection, have been proposed.
%They can be divided into ``classical'' and ``modern'' approaches. Among
%the classical methods one of the simplest and most well known is
%Forward
%Selection (Forward Selection). There are two common variants of
%Forward Selection. One version starts
%with a model containing no variables, then at each iteration adds the
%variable that produces the largest reduction in the sum of squares,
%conditional on the previously selected variables. The alternative
%implementation, and the version we concentrate on, also starts with a
%model containing no variables, then at each iteration adds the variable
%that has the highest correlation with the current residual vector,
%$\bY-X\hat\bbeta$, where the residuals are computed using an OLS fit
%on
%the already selected set of variables. Selecting the variable with
%highest
%correlation to the residual vector results in the greatest
%instantaneous
%reduction in the sum of squares for a small change in the regression
%coefficient so the two versions of Forward Selection can be seen as
%approximations to
%each other. While it is easy to construct theoretical examples where
%this
%greedy approach will fail, in practice it often gives good results.
%
More recently, interest has focused on an alternative class of penalization
methods, the most well known of which is the Lasso [\citet{tibshirani3}].
In addition to minimizing the usual sum of squares, the Lasso imposes an
$L_1$ penalty on the coefficients, which has the effect of automatically
performing variable selection by setting certain coefficients to zero and
shrinking the remainder. While the shrinkage approach can work well, it
has been shown that in sparse settings the Lasso often over-shrinks the
coefficients. Numerous alternatives and extensions have been suggested. A
few examples include SCAD [\citet{fan2}], the Elastic Net [\citet{zhu1}],
the Adaptive Lasso [\citet{zou1}], the Dantzig selector [\citet{candes1}], the
Relaxed Lasso [\citet{meinshausen2}], VISA [\citet{james12}] and the Double
Dantzig [\citet{james10}].

The Lasso has been made particularly appealing by the advent of the LARS
algorithm [\citet{efron3}] which provides a highly efficient means to
simultaneously produce the set of Lasso fits for all values of the tuning
parameter. The LARS algorithm starts with an empty set of variables and
then adds the predictor, say, $\bX_j$, most highly correlated with the
response. Next, the corresponding estimated coefficient, $\hat\beta
_j$, is
adjusted in the direction of the least squares solution. The algorithm
``breaks'' when the absolute correlation between $\bX_j$ and the residual
vector, $\bY-X\hat\bbeta$, is reached by the corresponding correlation
for another predictor. The new predictor, say, $\bX_k$, is then added to
the model, and the coefficients $\hat\beta_j$ and $\hat\beta_k$ are
increased toward their joint least squares solution until some other
variable's correlation matches those of $\bX_j$ and $\bX_k$, at which point
the new variable is also added to the model. This process continues until
all the correlations have reached zero, which corresponds to the ordinary
least squares solution.

By comparison, a common version of Forward Selection also starts with an
empty model and then iteratively adds to the model the variable most
highly correlated with the current residual vector. Next, the residuals are
recomputed using the ordinary least squares solution, based on the
currently selected variables. This algorithm repeats until all the variables
have been added to the model. In comparing Forward Selection with LARS,
one observes that the main difference is that the former method drives the
regression estimates for the currently selected variables all the way to
the least squares solution, while LARS only moves them part way in this
direction. Hence, the Lasso estimates the residual vector using shrunk
regression coefficients, while Forward Selection uses unshrunk estimates.
Which approach is superior? In Section \ref{methodsec} we show that, even
for toy examples with no noise in the response, neither universally
dominates the other. In some situations the Lasso's high level of
shrinkage produces the best results, while in other cases unshrunk
estimates work better.

In this paper we suggest viewing the Lasso and Forward Selection as two
extremes on a continuum of possible model selection rules. Instead of
selecting candidate models using either highly shrunk or else completely
unshrunk coefficients, we propose a methodology that can adaptively adjust
the level of shrinkage at each step in the algorithm. We call our approach
``Forward-Lasso Adaptive SHrinkage'' (FLASH). As with LARS, our
algorithm selects
the variable most highly correlated with the residuals and drives the
selected coefficients toward the least squares solution. However, instead
of stopping at the highly shrunk Lasso point or the zero shrinkage Forward
Selection point, FLASH uses the data to adaptively choose, at each
step, the optimal level of shrinkage before selecting the next variable.
FLASH includes Forward Selection and the Lasso as special cases, yet has
the same order of computational cost as the Lasso. After introducing FLASH
in the linear regression setting, we then extend it to the Generalized
Linear Models (GLM) domain. Thus, FLASH can also be used to perform
variable selection in high dimensional classification problems using, for
example, a logistic regression framework. This significantly expands the
range of problems that FLASH can be applied to. We show through
extensive simulation studies, as well as theoretical arguments, that FLASH
significantly outperforms Forward Selection, the Lasso and many
alternative methods, in both the regression and the GLM domains.

Our paper is structured as follows. In Section \ref{methodsec} we
demonstrate that neither Forward Selection nor the Lasso universally
dominate each other. We present the FLASH methodology in the linear
regression setting and outline an algorithm for efficiently constructing
its path. Some theoretical properties of FLASH are also discussed. Then
in Section \ref{simsec} we present a detailed simulation study to examine
the practical performance of FLASH in comparison to Forward Selection, the
Lasso and other competing methods. FLASH is extended to the GLM setting
in Section~\ref{glmsec} and further simulation results are provided. In
Section \ref{empsec} FLASH is demonstrated on several real world data
sets, predicting baseball salaries, real estate prices and whether an
internet image is an advertisement. These data sets all have many
predictors, up to $p=1430$, and involve both linear regression and GLM
scenarios. We end with a discussion in Section \ref{discsec}.

%s2 ###
\section{Methodology}
\label{methodsec}

Using suitable location and scale transformations, we can standardize the
data so that the response, $\bY$, and each predictor, $\bX_j$, are mean
zero with $\|\bX_j\|=1$. Throughout the paper we assume that this
standardization holds. However, all numerical results are presented on the
original scale of the data.

%s2.1 ###
\subsection{Lasso versus Forward Selection}
\label{L.vs.F}
As discussed in the introduction, both the LARS implementation of the Lasso
and the Forward Selection algorithm choose the variable with the
highest absolute correlation and then drive the selected regression
coefficients toward the least squares solution. The key difference is
that Forward Selection produces unshrunk estimates by utilizing the least
squares solution while the Lasso uses shrunk estimates by only driving the
coefficients part way. Which approach works better? It is not hard to
show that even in simple settings neither approach dominates the other.

%f1 ###
\begin{figure}

\includegraphics{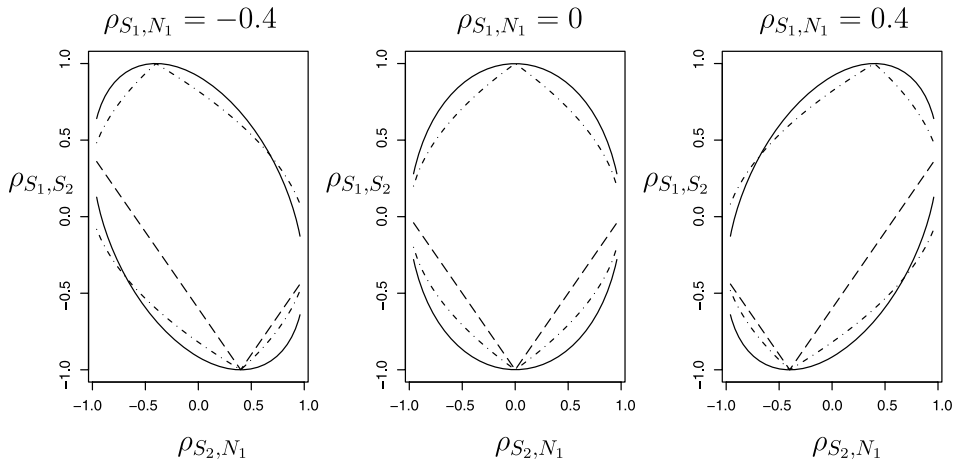}

\caption{Plots showing regions where the Lasso and
Forward Selection will identify the correct model for different
correlation structures. Points above the dashed lines correspond to
the Lasso regions. Points between the dash dot lines correspond to
Forward Selection. The solid lines provide the regions of feasible
correlation combinations.}\label{twodplots}
\end{figure}

Consider, for example, a scenario involving a linear model with two signal
predictors, one noise variable and no error term. Denote
by $\rho_{S_1,S_2}$ the correlation between the signal predictors and let
$\rho_{S_i,N_j}$ denote the correlation between the $i$th signal and $j$th
noise variable. Provided the coefficient for the first signal variable is
large enough, this variable is the one most highly correlated with the
response, thus
it is the first selected by both the Lasso and Forward Selection.
In this setting one can directly calculate the values of $\rho_{S_1,S_2},
\rho_{S_1,N_1}$ and $\rho_{S_2,N_1}$ where the Lasso or Forward Selection
selects the ``correct'' set of variables. Figure \ref{twodplots} provides
an illustration for three different values of $\rho_{S_1,N_1}$. The
regions between the dash dot curves correspond to the values of
$\rho_{S_1,S_2}$ and $\rho_{S_2,N_1}$ where Forward Selection will
identify the correct model. Alternatively, the regions above the
dashed curve represent the same situations for the Lasso. The solid
lines encompass the regions of feasible correlation combinations. Even in
this simplified example it is clear that there are many cases where
Forward Selection succeeds and the Lasso fails, and vice versa.

%f2 ###
\begin{figure}

\includegraphics{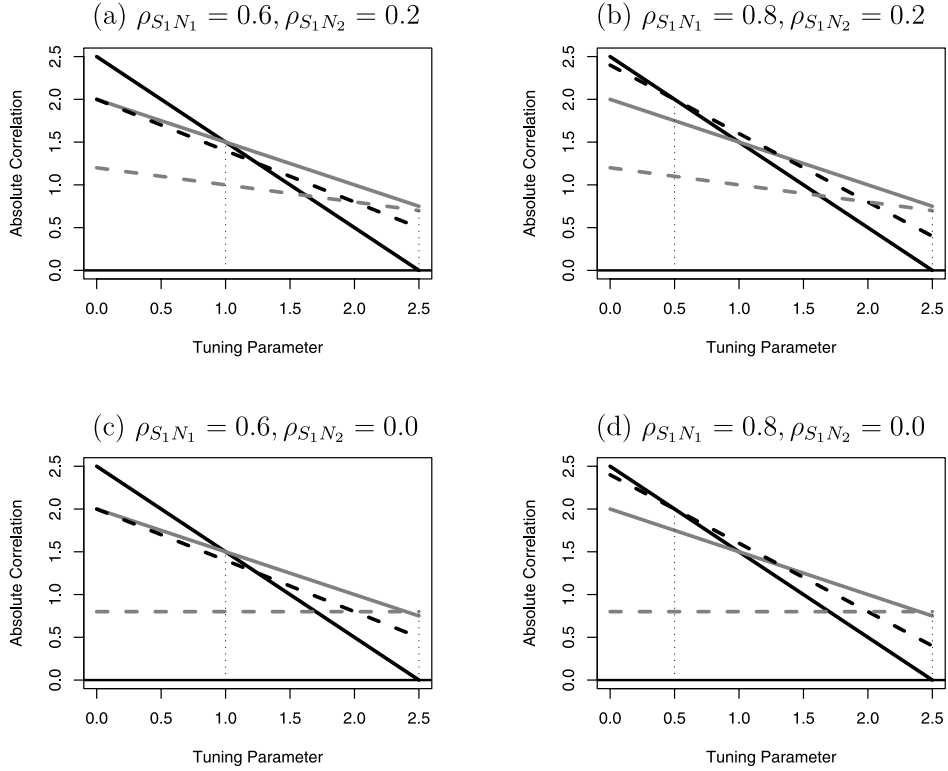}

\caption{Absolute correlations of the two signal
variables (black and gray solid) and two noise variables (black and gray
dashed) for different values of $\rho_{S_1N_1}$ and $\rho_{S_1N_2}$
in the example considered in Section \protect\ref{L.vs.F}. The first dotted
vertical indicates the Lasso break point, and the second dotted
vertical corresponds to Forward Selection. The line (other than black
solid) with the highest value at the break point indicates the variable
selected by the corresponding method. The Lasso succeeds only in \textup{(a)}
and \textup{(c)}, and Forward only in \textup{(a)} and \textup{(b)}.}\label{theoryplot3}
\end{figure}

Figure \ref{theoryplot3} graphically illustrates how the Lasso, Forward
Selection or both methods could fail, using the same simple setup with
one additional noise variable. For each plot the four lines
represent the absolute correlation between the corresponding variable and
the residual vector; solid lines for signal variables and dashed lines for
noise variables. The left-hand side of the plot corresponds to the null
model with all coefficients set to zero, and the lines show how the
correlations change as coefficients are adjusted toward the least squares
solution. Each plot represents different values of $\rho_{S_1,N_1}$ and
$\rho_{S_1,N_2}$. The values for the other
relevant parameters are fixed for all four plots at $\beta_1=2, \beta_2=1,
\rho_{S_1,S_2}=0.5, \rho_{S_2,N_1}=\rho_{S_2,N_2}=0.8$.

In all four plots the black solid line, representing the first signal
variable, has the maximal correlation for the null model, so both the
Lasso and
Forward Selection choose this variable first and drive its coefficient
toward the least squares solution. However, the Lasso stops when the
black line intersects with one of the other variables and adds that
variable next, the first vertical dotted line in each plot, while Forward
Selection drives the black line to zero, that is, the least squares solution,
and then selects the variable with the maximal correlation, the second dotted
line. For a method to choose the correct model it must select the second
signal variable, represented by the gray solid line. In
Figure \ref{theoryplot3}(a) the gray solid line is the highest at both
the Lasso and
Forward Selection stopping points, so both methods choose the correct
model. However, in Figure \ref{theoryplot3}(b) the Lasso selects the black
dashed noise variable, while Forward Selection still chooses the correct
model. Alternatively, in Figure \ref{theoryplot3}(c) the Lasso correctly
selects the gray signal variable, while Forward Selection chooses the gray
dashed noise variable. Finally, in Figure \ref{theoryplot3}(d) both the
Lasso and Forward Selection incorrectly select noise variables.

%s2.2 ###
\subsection{An adaptive shrinkage methodology}
\label{ad.shr.meth}

A key observation from Figure \ref{theoryplot3} is that in all four plots
the correct solid grey signal variable has the maximal correlation for at
least some levels of shrinkage, even in situations where the Lasso and
Forward Selection fail to identify the correct model. This example
illustrates that choosing the variable most highly correlated with the
residuals can work well provided the correct level of shrinkage is used.
This observation motivates our ``Forward-Lasso Adaptive SHrinkage''
(FLASH) methodology.

Like the Lasso and Forward Selection, FLASH begins with the null model
containing no variables and then implements the following procedure:
\begin{enumerate}
\item At each step add to the model the variable most highly correlated
with the
current residual vector.
\item Move the coefficients for the currently selected variables a given
distance in the direction toward the corresponding ordinary least
squares solution.
\item Repeat steps 1 and 2 until all variables have been added to the model.
\end{enumerate}
The FLASH algorithm is similar to that for LARS and Forward Selection. The
main difference revolves around the distance that the coefficients are
driven toward the least squares solution. For the $l$th step in the FLASH
algorithm this distance is determined by a tuning parameter, $\delta_l$.
Setting $\delta_l=0$ corresponds to the Lasso stopping rule, that is, driving
the coefficients until the maximum of their absolute correlations
intersect with that of another
variable. Alternatively, $\delta_l=1$ corresponds to the Forward Selection
approach where the coefficients are set equal to the corresponding
least squares
solution. However, setting $\delta_l=\frac{1}{2}$, for example,
causes the
coefficients to be driven half way between the Lasso and the Forward
Selection stopping points. As a result, FLASH can adjust the level of
shrinkage not just on the final model coefficients, as used previously
in, for example, the Relaxed Lasso, but also at each step during the
selection of
potential candidate models.

%f3 ###
\begin{figure}[b]

\includegraphics{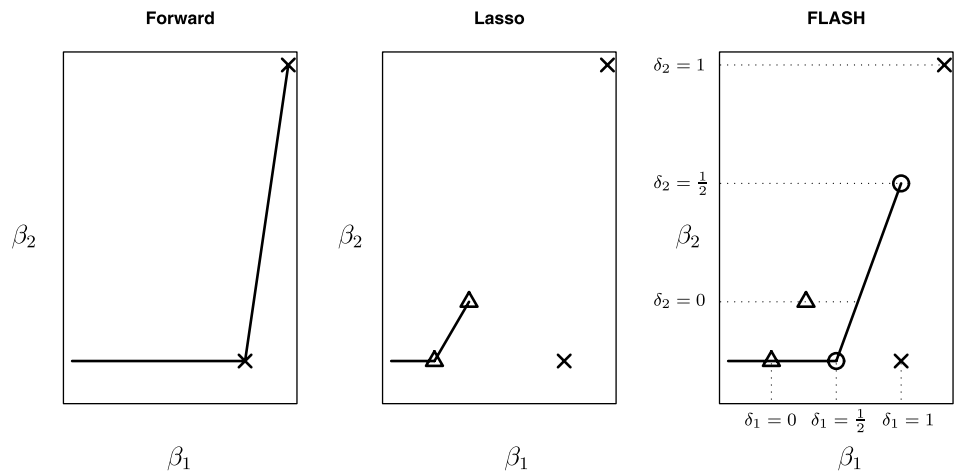}

\caption{Example coefficient paths for a two variable
example using Forward Selection (crosses), the Lasso (triangles) and
FLASH (circles).}\label{exampleplot}
\end{figure}

Figure \ref{exampleplot} illustrates potential coefficient paths, for the
first two variables selected, for each of the three different approaches.
The horizontal solid line in each plot shows the path for the first
variable selected, $\beta_1$. The first plot illustrates Forward
Selection where $\beta_1$ is driven all the way to the least squares
solution, represented by the first cross. Alternatively, the Lasso
(second plot) only drives $\beta_1$ a quarter of the way to the least
squares solution. Finally, the third plot shows one possible FLASH
solution. Here we have marked $\delta_1=0$ for the Lasso solution and
$\delta_1=1$ for the Forward Selection estimate. In this case we set
$\delta_1=\frac{1}{2}$ and, hence, the corresponding FLASH estimate for
$\beta_1$ is half way between the Lasso and Forward Selection
coefficients. The sloped solid line on each plot illustrates the
continuation of the paths to estimate both $\beta_1$ and $\beta_2$. Again,
Forward Selection drives $\beta_1$ and $\beta_2$ to their joint least
squares solution, while the Lasso estimate only moves part way in this
direction. The final plot shows the FLASH estimate, again setting
$\delta_2=\frac{1}{2}$.

In the following section we describe two different approaches for letting
the data select the optimal level of shrinkage at each step. In some
situations, for example, where a subset of the true variables has a high
signal, we may wish to adopt the Forward Selection approach with no
shrinkage. In other situations, for example, where there is a lot of noise,
the highly shrunk Lasso estimates may be preferred. But, as we show in
the simulation results, often a level of shrinkage between these two
extremes gives superior results. Another strength of FLASH is that its
coefficient path can be efficiently computed using a variant of the LARS
algorithm, which we outline next.

We use index $l$ to denote each step of the algorithm, but for
simplicity of the notation we omit this index wherever the meaning is
clear without it. Throughout the algorithm index set $\A$ represents the
correlations that are being driven toward zero, vector $\bc_{\A}$
contains the values of these correlations, and $X_{\A}$ denotes the matrix
consisting of the columns of $X$ associated with the set $\A$. We refer
to this set and the corresponding correlations as ``active.'' Note that
the active absolute correlations are driven toward zero at rates that are
proportional to their magnitudes:
\begin{enumerate}
\item Initialize $\bbeta^1=\bolds{0}$, $\A=\varnothing$ and $l=1$.
\item Update the active set $\A$ by including the index of the (new)
maximal absolute correlation. Compute the $|\A|$-dimensional direction
vector $\bh_{\A}=
(X_{\A}^TX_{\A} )^{-1} \bc_{\A}$. Let $\bh$ be the $p$-dimensional
vector with the components corresponding to $\A$ given by $\bh_\A$, and
the remainder set to zero.
\item Compute $\gamma_L$, the Lasso distance to travel in direction
$\bh$
until a new absolute correlation is maximal. We provide the formulas in
the \hyperref[app]{Appendix}, where we also show that $\gamma_F$, the Forward Selection
distance to travel in direction $\bh$ until the active correlations
reach zero, equals one.
Define $\gamma=\gamma_L+\delta_l(1-\gamma_L)$ and
let $\bbeta^{l+1} = \bbeta^l+\gamma\bh$. Set $l\leftarrow l+1$.
\item Repeat steps 2 and 3 until all correlations are at zero.
\end{enumerate}

Our attention has recently been drawn to the Forward Iterative Regression
and Shrinkage Technique (FIRST) in \citet{hwang1}, which can perform
effectively in sparse high-dimensional settings. FIRST also utilizes
aspects of the Forward Selection and Lasso approaches, but in a rather
different fashion than FLASH. For example, in the orthogonal design matrix
situation FIRST, when run to convergence, returns the Lasso fit, while
FLASH still produces a continuum of solutions between those of the Lasso
and Forward Selection.

%It exploits the fact
%that the LASSO has closed form solution when the predictor is
%one-dimensional. The explicit formula for this solution is repeatedly
%used in an iterative fashion to build the model until convergence
%occurs.
%However, this method is fundamentally different from ours, as it only
%updates one coefficient at each step, unshrinking it all the way to the
%level specified by the Lasso tuning parameter, while FLASH, similar to
%the
%Lasso, repeatedly unshrinks the coefficients for all the variables
%currently in the model.

%s2.3 ###
\subsection{Modifications to the algorithm}
\label{modif.sec}

In practice, we propose implementing FLASH with the following two
modifications. First, note that when all $\delta_l$ are set to zero,
the algorithm above reduces to the basic LARS algorithm, which does not
necessarily recover the Lasso path. To ensure that FLASH is a
generalization of the Lasso, we implement FLASH using the same
modification as the LARS algorithm uses to compute the Lasso path, that is,
if at any point on the path a coefficient hits zero, then the
corresponding variable is removed from the active set. A detailed description
of this modification is given in the \hyperref[app]{Appendix}.

Second, to account for the potential over-shrinkage of the coefficients in
a sparsely estimated model, we implement a ``relaxed'' version of FLASH,
which extends FLASH analogously to the way that the Relaxed Lasso extends
the Lasso. We unshrink each solution located at a breakpoint of the FLASH
path, connecting it via a path with the ordinary least squares solution on
the corresponding set of variables. We do this as soon as the FLASH
breakpoint is computed, in other words, right after the third step of the
algorithm. As with the Relaxed Lasso, the calculation of the corresponding
relaxation direction comes at no computational cost, as it coincides with
the current direction of the FLASH path. More specifically, the original
FLASH solution after step 3 is given by $\bbeta^l+\gamma\bh$, and the
corresponding OLS solution is given by $\bbeta^l+\bh$. The corresponding
relaxation path is given by linear interpolation between these two points.

For the remainder of this paper, when we refer to FLASH, we mean the
modified version.
In our numerical examples the final solution is selected via
cross-validation as a point on one of the relaxation paths, where each
of these continuous paths is replaced by its values on a fixed grid.

%The computation of $\gamma$ depends on the chosen values for $\delta_1,
%identical to the LARS step size and the FLASH and LARS algorithms
%become
%identical. However, for non-zero values of $\delta$ the FLASH step size
%becomes larger that that for LARS. The explicit formula for $\gamma$ is
%provided in the appendix.

%s2.4 ###
\subsection{Selection of tuning parameters}
\label{tuning.sec}

An important component of FLASH is the selection of the $\delta_l$
parameters. Clearly, treating each $\delta_l$ as an independent tuning
parameter is not feasible. Many model selection approaches could be
utilized. In this paper we investigate two possible approaches. The first,
``global FLASH,'' involves selecting a single value, $\delta$, for all the
step sizes, that is, assuming a common level of shrinkage throughout
the steps
of the FLASH algorithm. Hence, $\delta=0$ corresponds to the Lasso and
$\delta=1$ to Forward Selection. Using this approach, we first choose a
grid of $\delta$'s between $0$ and $1$ and then select the value giving
the lowest residual sum of squares on a validation data set or,
alternatively, the lowest cross-validated error. Global FLASH has the
advantage of only needing to select one $\delta$, which improves its
computational efficiency.

The second approach, ``block FLASH,'' allows for different values among
the $\delta_l$'s. However, to make the problem computationally feasible,
we constrain each $\delta_l$ to be either zero or one. The version of
block FLASH we focus on exclusively for the remainder of the paper
involves selecting a single ``break point'' with $\delta_{l^*}=1$ and
setting all remaining $\delta_l$'s to zero. This has the effect of
dividing FLASH into two stages. In the first stage a series of Lasso steps
(i.e., $\delta_l=0$) are performed to select the initial variables. At the
end of the first stage a Forward step (i.e., $\delta_{l^*}=1$) is
performed which has the effect of removing the coefficient shrinkage on
the currently selected variables. In the second stage further variables
are selected by performing a series of Lasso steps. As with global FLASH,
block FLASH has the advantage of only needing to select one tuning
parameter, the break point. In Section \ref{simsec} we provide simulation
results for both versions of FLASH. In practice, the two methods appear to
perform similarly. However, as illustrated below, we are able to
establish some
interesting theoretical properties for block FLASH.

Note that for each fixed $\delta$ or, correspondingly, each fixed $l^*$,
global and block FLASH both have the same computational cost as the LARS
algorithm.
%The same amount of
%computational effort is also required by the block FLASH procedure for
%each given location of the break point.
Because LARS is extremely efficient, so are the FLASH algorithms,
in particular, they require the same order of calculations as LARS if the
grid size for $\delta$ and the number of locations for $l^*$ are finite.
We propose using a five value grid for $\delta$, which worked
very well in our simulation study. The upper bound on the number of potential
locations for $l^*$ can be chosen based on the computational complexity of
the problem. Remember that $l^*$ represents the number of easily
identifiable predictors, so one might reasonably expect a relatively low
value.

%s2.5 ###
\subsection{Theoretical arguments}
\label{theorysec}
In this section we present some variable selection properties of FLASH, in
particular, conditions under which it can be shown to outperform the
Lasso. Throughout this section ``probability tending to one'' refers to
the scenario of $p$ going to infinity. For the standard case of
bounded $p,$ we could think of $n$ going to infinity instead, although some
minor modifications would need to be made to the statements of the
results. Let $K$ index the nonzero coefficients of $\bbeta$. We will
say that an estimator $\hbeta$ recovers the correct
signed support of $\bbeta$ if $\operatorname{sign}(\hbeta)=\operatorname{sign}(\bbeta)$,
where the equality is understood componentwise.

We will take a common approach of imposing bounds on the
maximum absolute correlation between two predictors. Define $S$ as
the number of signal variables, $\mu=\max_{j>k}|\bX_j^T\bX_k|$ and
let $\xi$ be an arbitrarily small
positive constant. The results of \citet{zhao2} and \citet{wain1} imply
that if
%
%e2 ###
\begin{equation}
\label{sml.coef.cnd} \min_{j\in K}|\beta_j|>c_1\sqrt{S\log p}
\end{equation}
and $\mu<\mu_{\mathrm{L}}(1-\xi)$ with $\mu_{\mathrm{L}}=1/[2S-1]$, then the Lasso solution
corresponding to an appropriate choice of the tuning parameter will
recover the correct signed support of $\bbeta$ with probability
tending to
one. Here the constant $c_1$ does not change with $n$ and $p$, and its
value is provided in the supplemental article [\citet{james17}]. On the
other hand, the number of true nonzero coefficients, $S$, is allowed to
grow together with $n$ and $p$.
Note that condition (\ref{sml.coef.cnd}) is stated for the rescaled
coefficients that correspond to the standardized predictor vectors.
On the original scale the right-hand side in (\ref{sml.coef.cnd})
would be of order $\sqrt{(S\log p)/n}$. Suppose, for example, that $S$
is bounded and $p$ grows polynomially in $n$. In
this case the lower bound on the magnitudes of the nonzero
coefficients, expressed on the original scale, goes to zero at the rate
$\sqrt{(\log n)/n}$.

%This result follows directly from Wainwright's Theorem 1 applied
%together with Zhao and Yu's
%Corollary 2 .

The correlation bound above is tight in the sense that for
each $\mu\ge\mu_L$ there are values of $X^TX$ and sign$(\bbeta)$ such
that the Lasso fails to recover the correct signed support. In the
following claim we identify a class of
such counterexamples.
\begin{claim}
\label{counter.ex} Let $\rho$ be a constant satisfying $\rho\ge\mu
_L$ and
let $j$ be an arbitrary index in $K^c$. Suppose that all the pairwise
correlations among the predictors indexed by $K\cup\{j\}$ equal $-\rho$,
and all the signs of the nonzero coefficients of $\bbeta$ are negative.
Then, with probability at least $1/2$, no Lasso solution recovers the
correct signed support of $\bbeta$.
\end{claim}

The proof of the claim is provided in the supplemental article [\citet{james17}]. Note that
for $\rho<1/S$ the correlation matrix can be easily made positive definite
by setting all the nonspecified pairwise correlations to zero.

Our Theorem \ref{vstheorem3} establishes that, under an additional
assumption on the magnitudes of the nonzero coefficients, block FLASH can
work in the situations where the Lasso fails. The intuition behind this
additional assumption is that for many regression problems there will be
some signal variables that are relatively easy to identify, while the
remainder pose more difficulties. The block FLASH procedure utilizes the
first group of signal variables in a more efficient fashion and hence is
better able to identify the remaining predictors. To mathematically
quantify this intuition, suppose that there exist indexes $a$ and $b$, such
that the corresponding true coefficients are nonzero and have a
significant separation in the magnitudes, that is, a large value
of $|\beta_a/\beta_b|$. We will refer to the coefficients
$\{\beta_j\dvtx |\beta_j|\ge|\beta_a|\}$ as large, and the coefficients
$\{\beta_j\dvtx 0<|\beta_j|\le|\beta_b|\}$ as small.
Theorem \ref{vstheorem3} below states that if the
ratio $|\beta_a/\beta_b|$ is sufficiently large, then the block FLASH
procedure will correctly identify the signal variables under a weaker
assumption on the maximal pairwise correlation. More specifically, at the
first stage the procedure will identify all the large nonzero coefficients
and not pick up any noise, and at the second stage it will pick up the
remaining nonzero coefficients without bringing in the noise.
%consider the situation where the non-zero coefficients of $\bbeta$
%split
%into three groups: coefficients that are the largest in magnitude,
%coefficients that are the smallest in magnitude, and the remaining
%coefficients. Denote the smallest absolute value in the first group by
%$m_1$ and the largest absolute value in the second group by $M_2$.
%Theorem \ref{vstheorem3} below states that if the ratio of $m_1$ to
%$M_2$
%is large enough, then block FLASH will correctly identify the signal
%variables under a weaker assumption on the maximal pairwise
%correlation.
As we discuss at the end of the section, the new correlation
bound, $\mu_{\mathrm{FL}}$, is strictly larger than the Lasso bound, $\mu_{\mathrm{L}}$.

\begin{theorem}
\label{vstheorem3} Suppose that condition (\ref{sml.coef.cnd}) holds,
inequality $|\beta_a/\beta_b|>c_3\sqrt{S}$ is satisfied for an
arbitrary pair of true nonzero coefficients,
and $\mu<\mu_{\mathrm{FL}}(1-\xi)$ for some arbitrary constant $\xi$.
Then, with an appropriate choice of the tuning parameters, the block
FLASH estimator recovers the correct signed support of $\bbeta$ with
probability tending to one.
\end{theorem}

%arbitrary constant $\xi$, condition (\ref{sml.coef.cnd}) holds
%and $m_1/M_2>c_3\sqrt{S}$. Then, with an appropriate choice of the
%tuning parameters, the block
%FLASH estimator recovers the correct signed support of $\bbeta$ with
%probability tending to one.
Here the constant $c_3$ does not change with $n$ and $p$, and its value is
provided in the supplemental article [\citet{james17}] together with the
proof of the theorem. Like the corresponding Lasso result in
\citet{wain1}, our theorem can handle subgaussian errors, that is, the
tails of
the error distribution are required to decay at least as fast as those of
a gaussian distribution. Relative to the Lasso result, the only new
assumption is on the separation between the large and the small
coefficients. Consequently, we are able to relax the requirement on the
pairwise correlations. According to Claim \ref{counter.ex}, the new
assumption does not allow us to relax the pairwise correlation requirement
for the Lasso, as the nonzero coefficients of $\bbeta$ affect the claim
only through their signs. Applying Theorem \ref{vstheorem3} in the setup
of the claim yields that the correct signed support of $\bbeta$ can be
recovered for all $\rho<\mu_{\mathrm{FL}}$. In other words, under an additional
assumption on the magnitudes of the nonzero coefficients, block FLASH
succeeds for $\rho\in[\mu_{\mathrm{L}},\mu_{\mathrm{FL}})$, where the Lasso fails.
%More specifically, bound (\ref{vs.th2.B2}) can be used for small
%values of $q_1$.
%
%The Lasso solution to which we refer corresponds to objective
%function (\ref{weight.lasso}) with the unit weights.
%
%The condition on the
%correlation matrix in Theorem \ref{vstheorem3} is optimized for $q
%2/3$. For smaller values of $q$ the bound is the same as that in
%Theorem \ref{vstheorem2}.
%
%Analysis of the proof of Theorem \ref{vstheorem3} shows that the bound
%$1/(2qS-1)$ needs to be imposed
%on the correlations that involve the variables in the first block to
%guarantee, assuming the ratio $m_1/M_2$ is
%sufficiently large, that the Lasso will correctly identify the first
%block of predictors. Note that the Lasso still needs
%to impose the bound $1/(2S-1)$ on the remaining correlations to ensure
%it recovers $\bbeta$ exactly, while block FLASH only needs to impose a
%weaker bound $1/[(2-q)S]$.

The correlation bound in Theorem \ref{vstheorem3} can be taken as
%
%e3 ###
\begin{equation}
\label{mu.fl}
\mu_{\mathrm{FL}}=\min\biggl\{\frac1{2(1-q_2)S-1} , \frac1{(2-q_1)S} \biggr\}.
\end{equation}
Here $q_1$ and $q_2$ are the fractions of large and small coefficients,
respectively, among all the nonzero coefficients. More specifically,
$q_1= |\{\beta_j\dvtx |\beta_j|\ge|\beta_a|\} |/S$
and $q_2= |\{\beta_j\dvtx 0<|\beta_j|\le|\beta_b|\} |/S$.
Observe that $\mu_{\mathrm{FL}}>\mu_{\mathrm{L}}$ when $q_1S>1$. In fact, the proof of
Theorem \ref{vstheorem3} reveals that the best possible
value of $\mu_{\mathrm{FL}}$ is strictly above $\mu_{\mathrm{L}}$ for all positive $q_1$.

%s3 ###
\section{Simulation results}
\label{simsec}

In this section we present a detailed simulation study comparing FLASH to
five natural competing approaches. We implemented both the global
(FLASH$_{\mathrm{G}}$) and the block (FLASH$_{\mathrm{B}}$) versions of our method discussed in
Section \ref{tuning.sec}. The tuning parameter $\delta$ in FLASH$_{\mathrm{G}}$ was
selected from a grid of five possible values, $\{0, 0.25, 0.5, 0.75, 1\}$. We
also tried a $\{0,1\}$ grid corresponding to the Lasso and Forward
Selection, and a $\{0, 0.5, 1\}$ grid, but the results were inferior, so we
do not report them here.\looseness=1

We compared FLASH to VISA, the Relaxed Lasso (Relaxo), the Adaptive Lasso
(Adaptive), Forward Selection (Forward) and the Lasso.
The Adaptive Lasso involves a preliminary step where the weights are
typically chosen by performing a least squares fit to the data. This is
not feasible for $p>n$, so we selected the weights using either the simple
linear regression fits, as suggested in \citet{huang1}, or a ridge
regression fit, as suggested in \citet{zou1}. The ridged fits dominated so
we only report results for the latter method here.
%In addition to the
%path version of Forward Selection we also implemented a version
%without a
%path and a version with shrinkage, where each Forward solution is
%connected to the zero vector via a linear path. The three methods
%performed similarly, so we only report the results for the first one.

Our simulated data consisted of five parameters which we varied: the
number of variables ($p=100$ or $p=200$), the number of training
observations ($n=50, n=70$ or $n=100$), the correlations among the columns
of the design matrix ($\rho=0$ or $\rho=0.5$), the number of nonzero
regression coefficients ($S=10$ or $S=30$) and the standard deviation
among the coefficients ($\sigma_\beta=0.5, \sigma_\beta=0.7$ or
$\sigma_\beta=1$). We tested most combinations of the parameters and
report a representative sample of the results. The rows of the design
matrix were generated from a mean zero normal distribution with a
correlation matrix whose off-diagonal elements were equal to $\rho$. The
error terms were sampled from the standard normal distribution, while the
regression coefficients were generated from a mean zero normal with
variance $\sigma_\beta^2$. For each simulated data set we randomly
generated a validation data set with half as many observations as the
training data and selected the various tuning parameters for each method
as those that gave the lowest mean squared error between the response and
predictions on the validation data. In particular, both the relaxation
parameter and the number of steps in the algorithm for the FLASH methods
and the Relaxed Lasso were selected using a validation set. For each
method and simulation we computed three statistics, averaged over $200$
data sets: False Positive, the number of variables with zero coefficients
incorrectly included in the final model; False Negative, the number of
variables with nonzero coefficients left out of the model; and L2 square,
the squared $L_2$ distance between the estimated coefficients and the
truth. Table \ref{simtable} provides the results.

%t1 ###
\begin{table}
\tabcolsep=0pt
\caption{Simulation results for each method. L2 square denotes the
squared $L_2$ distance between the
estimated coefficients and the truth, averaged over the 200 simulated
data sets. For each simulation scenario we placed in bold the best L2
square value together with the
L2 square value for any other method that was not statistically worse
at the $5\%$ level of significance}\label{simtable}
\begin{tabular*}{\tablewidth}{@{\extracolsep{\fill}}lcd{2.3}d{2.3}d{2.3}d{1.3}d{2.3}cd{2.3}@{}}
\hline
\textbf{Simulation} & \textbf{Statistic} & \multicolumn{1}{c}{\textbf{FLASH}$_{\mathbf{G}}$}&
\multicolumn{1}{c}{\textbf{FLASH}$_{\mathbf{B}}$}& \multicolumn{1}{c}{\textbf{VISA}} & \multicolumn{1}{c}{\textbf{Relaxo}} & \multicolumn{1}{c}{\textbf{Adaptive}} & \textbf{Forward} &
\multicolumn{1}{c@{}}{\textbf{Lasso}}\\
\hline
$n=100$, $p=100$ & False-Pos& 1.92 & 3.32 & 3.23 & 3.7 & 9.9 & 1.11 &
18.68 \\
$S=10$, $\rho=0$& False-Neg& 2.12 & 1.89 & 2.26 & 2.26 & 1.84 & 2.33 &
1.27 \\
$\sigma_\beta=1$& L2-sq& \multicolumn{1}{c}{\hspace*{4.5pt}\textbf{0.249}} & \multicolumn{1}{c}{\hspace*{4.5pt}\textbf{0.249}} & 0.292 & 0.308 &
0.342 & \hspace*{4.5pt}\textbf{0.244} & 0.436 \\[2pt]
$n=100$, $p=200$ & False-Pos& 1.99 & 3.91 & 3.53 & 3.87 & 12.61 & 1.07 &
21.18 \\
$S=10$, $\rho=0$& False-Neg& 2.32 & 2.09 & 2.44 & 2.45 & 2.44 & 2.48 &
1.64 \\
$\sigma_\beta=1$& L2-sq& \multicolumn{1}{c}{\hspace*{4.5pt}\textbf{0.267}} & 0.286 & 0.353 & 0.366 & 0.524
& \hspace*{4.5pt}\textbf{0.266} & 0.606 \\[2pt]
$n=50$, $p=100$ & False-Pos& 2.65 & 6.17 & 4.88 & 5.1 & 10.39 & 1.71 &
15.41 \\
$S=10$, $\rho=0$& False-Neg& 3.3 & 2.9 & 3.38 & 3.4 & 3.08 & 3.79 & 2.42
\\
$\sigma_\beta=1$& L2-sq& \multicolumn{1}{c}{\hspace*{4.5pt}\textbf{0.775}} & 0.848 & 0.996 & 1.021 & 1.228
& \hphantom{0}0.929 & 1.285 \\[2pt]
$n=50$, $p=200$ & False-Pos& 3.73 & 7.24 & 6.46 & 6.84 & 12.89 & 1.71 &
18.54 \\
$S=10$, $\rho=0$& False-Neg& 3.83 & 3.4 & 4.06 & 4.04 & 3.81 & 4.57 &
3.04 \\
$\sigma_\beta=1$& L2-sq& \multicolumn{1}{c}{\hspace*{4.5pt}\textbf{1.057}} & \multicolumn{1}{c}{\hspace*{4.5pt}\textbf{1.089}} & 1.477 & 1.496 &
1.999 & \hphantom{0}1.365 & 1.934 \\[4pt]
$n=100$, $p=100$ & False-Pos& 3.13 & 4.79 & 6.33 & 6.53 & 10.41 & 1.32 &
19.66 \\
$S=10$, $\rho=0.5$& False-Neg& 2.59 & 2.33 & 2.48 & 2.45 & 2.21 & 3.02 &
1.62 \\
$\sigma_\beta=1$& L2-sq& \multicolumn{1}{c}{\hspace*{4.5pt}\textbf{0.527}} & \multicolumn{1}{c}{\hspace*{4.5pt}\textbf{0.546}} & 0.629 & 0.656 &
0.661 & \hphantom{0}0.581 & 0.797 \\[2pt]
$n=100$, $p=200$ & False-Pos& 3.35 & 6.35 & 7.06 & 7.33 & 11.82 & 1.27 &
21.72\\
$S=10$, $\rho=0.5$& False-Neg& 3.12 & 2.88 & 3.06 & 3.11 & 3.01 & 3.57 &
2.23 \\
$\sigma_\beta=1$& L2-sq& \multicolumn{1}{c}{\hspace*{4.5pt}\textbf{0.608}} & 0.673 & 0.752 & 0.785 & 0.872
& \hphantom{0}0.655 & 1.029
\\[2pt]
$n=50$, $p=100$ & False-Pos& 5.12 & 8.31 & 7.23 & 7.44 & 11.27 & 2.42 &
16.2 \\
$S=10$, $\rho=0.5$& False-Neg& 3.95 & 3.38 & 3.79 & 3.88 & 3.53 & 4.82 &
2.99 \\
$\sigma_\beta=1$& L2-sq& \multicolumn{1}{c}{\hspace*{4.5pt}\textbf{1.732}} & \multicolumn{1}{c}{\hspace*{4.5pt}\textbf{1.743}} & 1.84 & 1.901 &
2.088 & 2.38 & 2.199\\[2pt]
$n=50$, $p=200$ & False-Pos& 5.82 & 9.62 & 8.8 & 8.77 & 12.91 & 2.37 &
18.28 \\
$S=10$, $\rho=0.5$& False-Neg& 5.14 & 4.45 & 5.04 & 5.12 & 4.9 & 6.25 &
4.34 \\
$\sigma_\beta=1$& L2-sq&\multicolumn{1}{c}{\hspace*{4.5pt}\textbf{2.399}} & \multicolumn{1}{c}{\textbf{2.35}} & 2.648 & 2.7 &
2.851 & \hphantom{0}3.094 & 2.934 \\[4pt]
$n=50$, $p=100$ & False-Pos& 10.6 & 13.73 & 11.34 & 12.09 & 15.62 & 4.39
& 17.16 \\
$S=30$, $\rho=0$& False-Neg& 14.23 & 12.1 & 14.12 & 13.89& 12.91 & 21.7
& 11.95 \\
$\sigma_\beta=1$& L2-sq& 10.559 & \multicolumn{1}{c}{\hspace*{4.5pt}\textbf{9.051}} & 10.749 & 10.743 &
11.132 & 19.792 & 11.316 \\[4pt]
$n=100$, $p=100$ & False-Pos& 3.54 & 4.72 & 6.09 & 6.19 & 10.52 & 1.84 &
17.86 \\
$S=10$, $\rho=0.5$& False-Neg& 3.73 & 3.54 & 3.51 & 3.56 & 3.34 & 4.24 &
2.46 \\
$\sigma_\beta=0.7$& L2-sq& \multicolumn{1}{c}{\hspace*{4.5pt}\textbf{0.625}} & \multicolumn{1}{c}{\hspace*{4.5pt}\textbf{0.624}} & 0.692 & 0.705
& 0.724 & \hphantom{0}0.707 & 0.787 \\[2pt]
$n=100$, $p=200$ & False-Pos& 4.06 & 6.21 & 7.11 & 7.48 & 11.69 & 1.68 &
21.46 \\
$S=10$, $\rho=0.5$& False-Neg& 4.05 & 3.81 & 3.87 & 3.93 & 3.7 & 4.68 & 3 \\
$\sigma_\beta=0.7$& L2-sq&\multicolumn{1}{c}{\hspace*{4.5pt}\textbf{0.686}} & 0.731 & 0.788 & 0.794 &
0.799 & 0.77 & 0.922 \\[4pt]
$n=100$, $p=100$ & False-Pos& 3.81 & 4.88 & 6.11 & 5.93 & 9.65 & 1.99 &
15.78 \\
$S=10$, $\rho=0.5$& False-Neg& 4.74 & 4.49 & 4.46 & 4.51 & 4.16 & 5.48 &
3.42 \\
$\sigma_\beta=0.5$& L2-sq& \multicolumn{1}{c}{\hspace*{4.5pt}\textbf{0.559}} & \multicolumn{1}{c}{\hspace*{4.5pt}\textbf{0.545}} & 0.576 & 0.584
& 0.596 & \hphantom{0}0.683 & 0.633 \\[2pt]
$n=100$, $p=200$ & False-Pos& 3.69 & 5.25 & 5.76 & 6.13 & 10.11 & 1.54 &
17.73 \\
$S=10$, $\rho=0.5$& False-Neg& 5.38 & 5.08 & 5.29 & 5.32 & 4.88 & 6.03 &
4.24 \\
$\sigma_\beta=0.5$& L2-sq& \multicolumn{1}{c}{\hspace*{4.5pt}\textbf{0.664}} & \multicolumn{1}{c}{\hspace*{4.5pt}\textbf{0.662}} & \multicolumn{1}{c}{\hspace*{4.5pt}\textbf{0.696}} &
0.709 & 0.715 & \hphantom{0}0.761 & 0.769\\
\hline
\end{tabular*}
\end{table}

The first four simulations corresponded to $\rho=0$, while the next
four were
generated using $\rho=0.5$. The ninth simulation was a denser
case with $S=30$. Finally, the last four simulations represent harder
problems with $\sigma_\beta=0.7$ or $0.5$, reducing the signal to noise
ratio from $10$ to $4.9$ and $2.5$, respectively. For the L2 square
statistic we performed tests of statistical
significance, comparing each method to the best FLASH approach. For each
simulation we placed in bold the L2 square value for the best method
and any
other method that was not statistically worse at the $5\%$ level of
significance. For example, in the first
simulation with $100$ variables and $100$ observations both versions of
FLASH and Forward were statistically indistinguishable from each
other. However, in the third simulation with $100$ variables and $50$
observations FLASH$_{\mathrm{G}}$ was statistically superior to all other
methods. Most of the standard errors for the L2 square statistic were
relatively low, approximately $4\%$ of the statistic's value. However, as
has been observed previously, we found that the Forward method often gave
more variable estimates than the other approaches, with some standard
errors as high as $8\%$ of the statistic's value.

None of the thirteen simulations contained a situation where one of the
competing methods was statistically superior to FLASH, while in ten of the
simulations FLASH was statistically superior to all other methods. In
general, Forward Selection performed well in the easiest scenarios with
large $n$, zero
correlation, $\rho$, and higher signal, $\sigma_\beta=1$. In particular,
Forward Selection performed very poorly in the denser $S=30$ scenario,
while this
was a favorable situation for the Lasso. FLASH was still superior to both
methods in this simulation setup. The Adaptive Lasso, VISA and Relaxo all
provided improvements over the Lasso, though the latter two methods
generated the largest increase in performance. The two versions of FLASH
performed at a similar level, though FLASH$_{\mathrm{G}}$ seemed slightly better in
the sparser cases, while FLASH$_{\mathrm{B}}$ was superior in the denser $S=30$
situation. FLASH$_{\mathrm{G}}$ also required less computational effort, because its
path only needed to be computed once for each of the five potential values
of $\delta$.

Overall, Forward Selection had low false positive but high false
negative rates. In
comparison to VISA and Relaxo, FLASH$_{\mathrm{G}}$ had the lowest false positive
rates and similar or lower false negative rates. Alternatively, FLASH$_{\mathrm{B}}$
had very low false negative rates and similar false positive rates.
Overall, FLASH selected sparser models than VISA, the Relaxed Lasso and
the Adaptive Lasso, and significantly sparser models than the Lasso.

%Figure \ref{boxplots} displays the results for the second and third
%simulation scenarios in the correlated case (each scenario has $100$
%variables, the left and right figures correspond to $70$ and $100$
%training observations, respectively). The box plots display the natural
%logs of the L2-sq values for FLASH$_5$, Relaxo, Visa, Forward
%Selection and Lasso. In
%the first scenario the performance of the Forward Selection method is
%highly variable,
%the results are extremely good for the ``easier'' data sets and
%extremely
%bad for the ``harder'' ones. FLASH seems to lean towards Forward
%Selection in the easier
%cases and Relaxo in the harder ones. As a result, FLASH is clearly the
%best performing method in this simulation setting. In the second
%scenario
%Forward Selection and FLASH show similar results and significantly
%outperform the
%competition.

%results for simulations two and three in the correlated case.
%The L2-sq values are plotted on the natural log scale.}

%s4 ###
\section{Extending to generalized linear models}
\label{glmsec}

%s4.1 ###
\subsection{Methodology}
In the generalized linear model framework for a response variable,
$Y$, with distribution
\[
p(y;\theta,\phi) = \exp\biggl(\frac{y\theta- b(\theta)}{a(\phi)} + c(y,
\phi) \biggr),
\]
one models the relationship between predictor and
response as $g(\mu_i) = \sum_{j=1}^pX_{ij}\beta_j$, where $\mu_i =
E(Y;\theta_i,\phi) = b'(\theta_i)$, and $g$ is referred to as the link
function. Common examples of $g$ include the identity link used for normal
response data and the logistic link used for binary response data. For
notational simplicity we will assume that $g$ is chosen as the canonical
link, though all the ideas generalize naturally to other link functions.
The coefficient vector $\bbeta$ is generally estimated by maximizing
the log
likelihood function,
%
%e4 ###
\begin{equation}
\label{standard.ml}
l(\bbeta) = \sum_{i=1}^n \bigl( Y_i\bX_i^T\bbeta-
b(\bX_i^T\bbeta) \bigr).
\end{equation}
However, when $p$ is large relative to $n$,
the maximum likelihood approach suffers from problems similar to those
of the
least squares approach in linear regression. First, maximizing
(\ref{standard.ml}) will not produce any coefficients that are exactly
zero, so no variable selection is performed. As a result, the final model
is less interpretable and probably less accurate. Second, for large $p$
the variance of the estimated coefficients will become large and when
$p>n$, function (\ref{standard.ml}) has no unique minimum.

Various solutions have been proposed. \citet{park1} discuss a natural GLM
extension of the Lasso (GLasso) where, for a fixed $\lambda$, they choose
$\bbeta$ to minimize
%
%e5 ###
\begin{equation}
\label{glasso}
l(\bbeta,\lambda)=-l(\bbeta)+\lambda\|\bbeta\|_1.
\end{equation}
The coefficient paths for the GLasso are not generally piecewise linear,
but Park and Hastie present an algorithm for approximating the true path.
Forward Selection can also be easily extended to the GLM domain by
starting with an empty set of variables and, at step $l$, adding to the
model the variable that maximizes the $j$th partial derivative of the log
likelihood, $l'_j(\hat\bbeta_l)$. One then sets $\hat\bbeta_{l+1}$ equal
to the maximum likelihood solution corresponding to the currently selected
variables and repeats. Note that $l'_j(\hat\bbeta_l)=\bX_j^T(\bY-\hat\mu)$,
which is just the correlation between the $j$th predictor and the
residuals. When using Gaussian errors with the identity link function,
$\hat\mu=X\bbeta$, so this algorithm reduces back to standard Forward
Selection in the regression setting.

%f4 ###
\begin{figure}[b]

\includegraphics{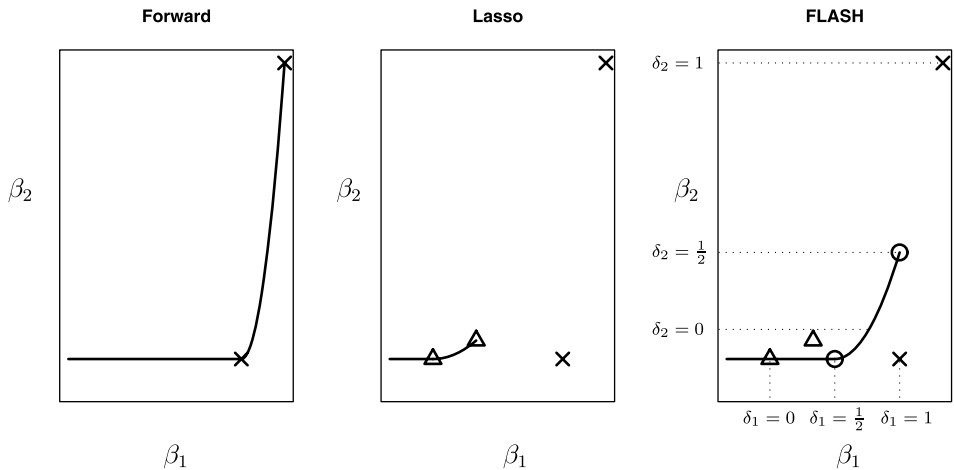}

\caption{Example coefficient paths for a two variable
GLM example using Forward Selection (crosses), the Lasso (triangles) and
FLASH (circles).}\label{glmplot}
\end{figure}

The GLM versions of the Lasso and Forward Selection also suggest a natural
extension of FLASH to this domain. In the GLM FLASH algorithm we again
start with an empty set of variables, $\A_1$, and $\hat\bbeta_1=\bolds{0}$.
Then at step $l$ we add to the model the variable that maximizes the $j$th
partial derivative of the log likelihood, $l'_j(\hat\bbeta_l)$, that
is, the
variable with maximal correlation. Finally, we drive $\hat\bbeta
_{l+1}$ in
the direction toward the maximum likelihood solution with the distance
determined by $\delta_{l}$. Again, $\delta_{l}=0$ corresponds to
shifting $\hat\bbeta_{l}$ as far as the Lasso stopping point, while
$\delta_{l}=1$ represents the maximum likelihood solution. However, one
key difference between the GLM and standard versions of FLASH is that,
because the coefficient paths are no longer piecewise linear, the
coefficients do not move in a linear fashion toward the maximum
likelihood solution.

Figure \ref{glmplot} provides a pictorial example in the same two variable
domain as for Figure \ref{exampleplot}. The GLasso still significantly
shrinks the coefficients relative to the Forward Selection approach.
Alternatively, FLASH provides an in between level of shrinkage. However,
notice that the coefficient paths now move in a curved fashion toward the
maximum likelihood solution. It is possible to compute this nonlinear
path on a grid of tuning parameters, and we present the precise algorithm
in the \hyperref[app]{Appendix}.

The block FLASH approach is particularly appealing in the GLM setting,
because it is both conceptually simple and easy to implement. With this
method FLASH follows the GLasso path for the first $l^*-1$ steps, that is,
$\delta_1=\delta_2=\cdots=\delta_{l^*-1}=0$. At this point the maximum
likelihood solution for the currently selected variables is computed,
that is,
$\delta_{l^*}=1$. Finally, the GLasso path is followed again with zero
penalty on the variables corresponding to $\A_{l^*}$, that is,
$\delta_{l^*+1}=\cdots=0$. We compute the GLM version of block FLASH using
two implementations of the R function $\operatorname{glmnet}(\cdot)$ [\citet{friedman8}],
which uses a coordinate
descent algorithm to minimize (\ref{glasso}). We first use $\operatorname{glmnet}(\cdot)$
to compute the path prior to $l^*$ and then make a
second call to the function to compute the path after $l^*$, placing zero
penalty on the variables selected in the first step.

%s4.2 ###
\subsection{Simulation study}
In this section we provide a simulation comparison of the block GLM FLASH
method with several other standard GLM approaches. In particular, we compared
FLASH to ``GLasso,'' ``GRelaxo,'' ``GForward'' and the standard ``GLM.'' GLasso
is implemented using the R function $\operatorname{glmnet}(\cdot)$. GRelaxo takes the
same sequence of models suggested by GLasso but unshrinks the final
coefficient estimates using a standard GLM fit to the nonzero
coefficients. GForward uses the approach outlined previously.

We simulated responses from the Bernoulli distribution using the
logistic link
function. The data were generated with $p=100$
variables, but we increased the sample size to $n=400$ as the Bernoulli response
provided less information compared to the Gaussian response. The correlation among the
predictors was set to either
$\rho=0$ or $\rho=0.5$, and the number of true signal variables was
set to
either $S=10$ or $S=15$. Finally, the nonzero regression coefficients were
randomly sampled from either a point mass distribution, with
probability a half of being $0.5$ or $-0.5$, or the standard normal
distribution. The tuning parameters for all methods were selected as those
that minimized the ``deviance'' on a validation data set with $n=200$
observations. In all other respects the simulation setup was the same as
the one we used in the linear regression setting.

%t2 ###
\begin{table}
\caption{Simulation results for each method using a Bernoulli
response. L2 square denotes the squared $L_2$ distance between the
estimated coefficients and the truth, averaged over the 200 simulated
data sets. For each simulation scenario we placed in bold the best L2
square value together with the
L2 square value for any other method that was not statistically worse
at the $5\%$ level of significance}
\label{simtable2}
\begin{tabular*}{\tablewidth}{@{\extracolsep{\fill}}lccd{2.3}d{2.3}d{1.3}d{5.3}@{}}
\hline
\textbf{Simulation} & \textbf{Statistic} & \textbf{FLASH}$_{\mathbf{B}}$&
\multicolumn{1}{c}{\textbf{GRelaxo}} & \multicolumn{1}{c}{\textbf{GLasso}} & \multicolumn{1}{c}{\textbf{GForward}} & \multicolumn{1}{c@{}}{\textbf{GLM}}\\
\hline
$n=400$, $p=100$ & False-Pos& 4.38\hphantom{0} & 4.47 & 16.61 & 1.18 & 90 \\
$S=10$, $\rho=0$& False-Neg& 0.34\hphantom{0} & 0.45 & 0.06 & 0.55 & 0 \\
$\beta=\pm0.5$& L2-sq& \textbf{0.461} & 0.488 & 0.71 & \multicolumn{1}{c}{\textbf{0.454}} &
6.065 \\[3pt]
$n=400$, $p=100$ & False-Pos& 9.1\hphantom{00} & 10.11 & 15.57 & 1.54 & 90 \\
$S=10$, $\rho=0.5$& False-Neg& 1.69\hphantom{0} & 1.82 & 0.92 & 3.51 & 0 \\
$\beta=\pm0.5$& L2-sq& \textbf{1.081} & 1.147 & \multicolumn{1}{c}{\hspace*{4.5pt}\textbf{1.107}} & 1.415 &
10.559 \\[6pt]
$n=400$, $p=100$ & False-Pos& 2.94\hphantom{0} & 3.22 & 17.88 & 0.6 & 90 \\
$S=10$, $\rho=0$& False-Neg& 2.94\hphantom{0} & 3.07 & 1.8 & 3.24 & 0 \\
$\beta=N(0,1)$& L2-sq& \textbf{0.72} & 0.779 & 1.954 & \multicolumn{1}{c}{\textbf{0.668}} &
99.794\\[3pt]
$n=400$, $p=100$ & False-Pos& 5.37\hphantom{0} & 7.63 & 17.41 & 0.74 & 90 \\
$S=10$, $\rho=0.5$& False-Neg& 3.21\hphantom{0} & 3.21 & 2.17 & 4.08 & 0 \\
$\beta=N(0,1)$& L2-sq& \textbf{1.168} & 1.392 & 1.967 & 1.289 & 58.08 \\[6pt]
$n=400$, $p=100$ & False-Pos& 4.42\hphantom{0} & 5.8 & 15.55 & 0.7 & 85 \\
$S=15$, $\rho=0.5$& False-Neg& 5.57\hphantom{0} & 5.38 & 3.66 & 6.81 & 0 \\
$\beta=N(0,1)$& L2-sq& \textbf{2.302} & 2.692 & 4.211 & 2.487 & 11903.88\\
\hline
\end{tabular*}
\end{table}

The results from five different simulations are provided in
Table \ref{simtable2}. Standard GLM performs very poorly. Note we have
reported the median errors for this method because the algorithm did not
converge properly for some simulations. GForward was competitive with
FLASH$_{\mathrm{B}}$ when using uncorrelated predictors but deteriorated in the
correlated situation. In all scenarios FLASH$_{\mathrm{B}}$ either had the lowest
L2-sq statistic or was not statistically different from the best. In the
last two simulations it was statistically superior to all the other
approaches.

%s5 ###
\section{Empirical analysis}
\label{empsec}

We implemented the global, block and GLM versions of FLASH on three
different real world data sets. The first contained salaries of professional
baseball players (obtained from StatLib, Department of Statistics, CMU).
For each player a number of statistics were recorded, such as career runs
batted in, walks, hits, at bats, etc. We then used these variables to
predict salaries. After including all possible interaction terms, the
data set contained $n=263$ observations and $p=153$ predictors.
%This
%represented a somewhat difficult regression problem, because of the
%large
%number of predictors relative to the number of observations.

\begin{figure}
\begin{tabular}{cc}

\includegraphics{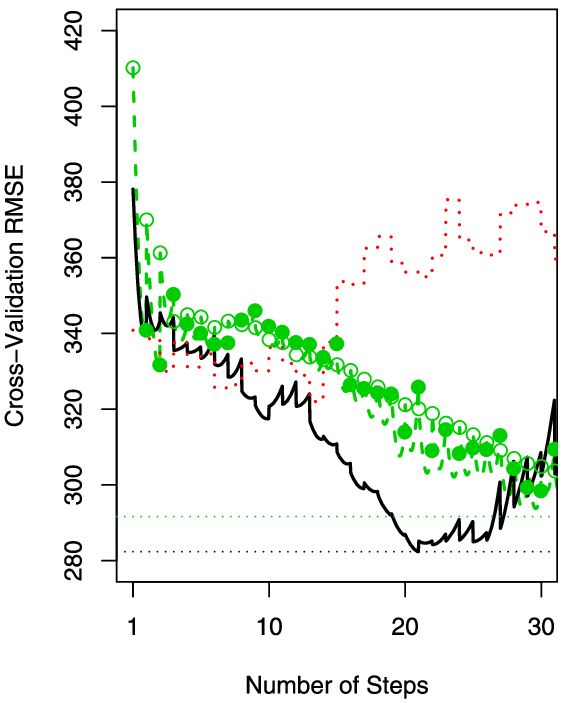}
&\includegraphics{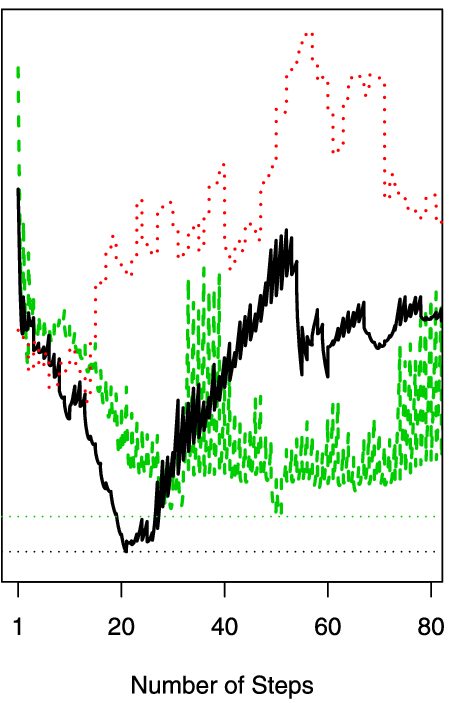}\\
(a)&(b)
\end{tabular}
\caption{\textup{(a)} The cross-validated root mean squared error
plotted versus the number of steps in the corresponding algorithm for
four different methods in the example of predicting baseball players'
salary. The methods displayed are Forward Selection (dotted line),
Lasso (open circles), OLS fits corresponding to the Lasso models
(solid circles), Relaxed Lasso (dashed line, connecting the
open and the solid circles) and
FLASH with $\delta=0.25$ (solid black line). \textup{(b)} Same as \textup{(a)} with more
steps.}\label{cverr}
\end{figure}

We tested three competitors to FLASH, namely, Lasso, Forward Selection and
the Relaxed Lasso. For each of the four methods ten-fold cross-validation
was used to compute the root mean squared error (RMSE) in prediction
accuracy at various points of the coefficient path. The final results are
illustrated in Figure~\ref{cverr}(a). The open circles represent the Lasso
RMSE's evaluated at the break points of the LARS algorithm.
Alternatively, the solid circles show the errors corresponding to the
least squares fits for the models selected by the Lasso. The dashed line
that connects the open and the solid circles illustrates the Relaxed Lasso
fit as the coefficient shrinkage is reduced from the Lasso estimate
(maximum shrinkage) to the least squares fit (no shrinkage). The dotted
line corresponds to Forward Selection.
%This is a step function because no
%shrinkage is used to estimate the coefficients so the fit remains
%constant
%within each step.
Finally, the black solid line represents the global FLASH fit
with $\delta=0.25$. We have fixed the value of the $\delta$ parameter
to ensure fair comparison with other methods on the basis of the
cross-validated RMSE. In our simulations we used a five point grid,
where the end points corresponded to the Relaxed Lasso and Forward
Selection, respectively. Among the remaining three values, we decided
to pick $\delta=0.25$, as it is closer to the Relaxed Lasso, which is
in general a more reliable method then Forward Selection.

%f5 ###

From step 15 onward, Forward Selection begins to significantly
deteriorate, while FLASH continues to improve and eventually achieves
the lowest
error rate of all four methods at approximately step 20.
Figure \ref{cverr}(b) plots the cross-validated error paths out to 80
steps. The Relaxed Lasso achieves its optimal results at approximately
step 50, which corresponds to a $34$ variable model. Not only is the
optimal error rate worse than that for FLASH, but the corresponding model
contains twice as many variables as the model selected by FLASH, which
only had $17$ variables. Our simulation results pointed to a
similar phenomenon, and we have noticed in other real data sets that FLASH
tends to select sparser models, suggesting FLASH may have an advantage in
terms of inference in addition to prediction accuracy.

The second data set we examined was the Boston Housing data, commonly used
to compare different regression methods. After including interaction
terms this data contained $90$ predictors of the average house value in
$506$ locations. To test the $p\approx n$ scenario, $90$ observations were
randomly sampled for the training data, $45$ observations for a validation
data set, and the remainder for the testing data. We implemented the block
FLASH approach and compared it to the Lasso, Relaxo and Forward Selection.
Least squares fits were used for the final coefficient estimates on all
methods except the Lasso. Hence, for example, the Relaxed Lasso
solutions simplify to the OLS solutions computed for the sequence
of models specified by the Lasso. Each approach was fitted using the
training data,
with the tuning parameters chosen using the validation data, and then the
mean squared error was computed on the test data. This procedure was
repeated using $100$ different random samplings of the data, to average
out any effect from the choice of test sets. Table \ref{bostontable}
shows, for each method, the average mean squared error as well as the
average number of coefficients chosen in the final model.
Block FLASH achieves the lowest mean squared error. In addition, FLASH,
Relaxo and Forward Selection all choose significantly smaller models than
the Lasso, making their results more interpretable. On average, block FLASH
selected $8.67$ variables before implementing the Forward step. FLASH
resulted in lower MSE than Relaxo in 62 random splits of the data,
the two methods had the same MSE in 3 splits, and FLASH had a higher
MSE in 35 of the splits. The corresponding numbers comparing FLASH to
the Lasso and FLASH to Forward Selection
were 63/0/37 and 81/0/19.

%
%t3 ###
\begin{table}
\tablewidth=250pt
\caption{Mean squared errors, averaged over $100$ test data sets,
and average number of variables selected, for the Boston Housing data}
\label{bostontable}
\begin{tabular*}{\tablewidth}{@{\extracolsep{\fill}}lcccc@{}}
\hline
&\textbf{FLASH} & \textbf{Relaxo} & \textbf{Lasso} & \textbf{Forward}\\
\hline
MSE & 27.01 & 28.30 & 29.56 & 33.03\\
Number of coefficients & 18.93 & 17.13 & 26.99 & 16.8\hphantom{0}\\
\hline
\end{tabular*}
\end{table}

The final data set that we examined was the internet advertising data
available at the UC-Irvine machine learning repository. The response was
categorical, indicating whether or not each image was an advertisement.
The predictors recorded the geometry of the image as well as whether
certain phrases occurred in and around the image \textsc{url}'s. After
preprocessing, the data set contained $n=2359$ observations and $p=1430$
variables. The large value of $p$ presented significant statistical and
computational difficulties for standard approaches, with the $\operatorname{glm}(\cdot)$
function in R taking almost fifteen minutes to run and producing NA
estimates for most coefficients. However, we were able to implement GLM
block FLASH, randomly assigning two-thirds of the observations to the
training data set and the remainder to the validation data set. FLASH
selected a twenty-seven variable model, with the five most important variables,
in terms of the order that they entered the model, being the width of the
image, whether the image's anchor \textsc{url} contained the phrase
``com,'' whether the \textsc{url} contained the phrase ``ads,''
and whether the anchor \textsc{url} contained the phrases ``click''
and ``adclick.'' We also fixed the tuning parameters at the
selected values and used a bootstrap analysis to produce pointwise
confidence intervals on the coefficients.
%used the tuning parameters selected
%by the validation set to conduct a bootstrap analysis on the original
%data.
GLM block FLASH ran very efficiently, taking approximately twenty
seconds to produce the corresponding estimator on each bootstrapped
data set. The misclassification error on the validation set was $2.9\%$
for FLASH, while it was $3.2\%$, $4.0\%$
and $5.0\%$, for GRelaxo, GLasso and GForward, respectively. The sizes
of the models selected by the last three methods were $38$, $77$ and
$16$, respectively.

%s6 ###
\section{Discussion}
\label{discsec}

The main difference between Forward Selection and the Lasso is in the
amount of shrinkage used to iteratively estimate the regression
coefficients. For any given data set there is no particular reason that
either the zero shrinkage of Forward Selection or the extreme shrinkage
of the Lasso must produce the best solution. FLASH allows the data to
dictate the optimal level of shrinkage at the model selection stage. This
is quite different from approaches such as Relaxo that adjust the
level of shrinkage after the model has been selected but not while choosing
the sequence of models to consider. As a result, FLASH often produces
sparser models with superior predictive accuracy.

Computational efficiency is always important for high-dimensional
problems. The standard FLASH algorithm is very similar to LARS and hence
involves a relatively small computational expense. In addition, the block
FLASH approach can easily be formulated as a penalized regression problem
with the usual $L_1$ penalty before the Forward step and zero penalty on
certain variables after this step. Hence, even more efficient methods,
such as the recent work on pathwise coordinate descent algorithms
[\citet{friedman7}], can be used to compute the path, not only for
regression problems, but also for our extension to GLM data. Indeed, the
$\operatorname{glmnet}(\cdot)$ function that we used to fit GLM block FLASH utilizes a
coordinate descent algorithm.

\begin{appendix}

%s7 ###
\section{Step 3 of the FLASH algorithm}\label{app}

Let $c_{i*}$ be one of the active correlations with the maximum absolute
value. Then, as with LARS, the first time a nonactive absolute
correlation reaches the ``active'' maximum corresponds to the step size of
\[
\gamma_L = \min^+_{j\in\A^c}
\biggl\{\frac{c_{i*}-c_j}{ (\bX_{i*}-\bX_j)^T X\bh},
\frac{c_{i*}+c_j}{ (\bX_{i*}+\bX_j)^T X\bh} \biggr\},
\]
where the minimum is taken over the positive components.

Along the direction $\bh$, all active correlations reach zero at the
same time. Hence, the Forward Selection step size is given by
\[
\gamma_{F} = \frac{c_{i*}}{ \bX_{i*}^T X\bh}=\frac{c_{i*}}{ \bX
_{i*}^T X_{\A}(X^T_{\A}X_{\A})^{-1}\bc_{\A}}=1.
\]

%and the first point that non-zero coefficient crosses zero is given
%by $\gamma_3=\min^+_j \{-\beta^l_j/h_j \}$. Combining the
%above with the possibility that $C=0$, we get
%$\gamma=\min\{\gamma_1,\gamma_2,\gamma_3,c_k/\bx^T_kX\bh\}$.

%s8 ###
\section{Zero crossing modification}
The basic FLASH algorithm described in Section \ref{ad.shr.meth}
shares the following property with the basic LARS algorithm: once a
variable enters the model, it does not leave. Recall that the Lasso
solution path can be obtained from the modified LARS algorithm,
where if a coefficient hits zero, the corresponding variable is
removed from the active set, and hence the model as well. When a
variable is removed, the corresponding absolute correlation goes
below the value it would be at if it remained active. The variable
rejoins the model if its absolute correlation reaches the value it
would be at, had the variable stayed in the model. We provide a similar
modification to the FLASH algorithm.

\begin{definition}[(Zero crossing modification)]
When a coefficient hits zero, the corresponding variable is removed from
the active set. The variable is added back to the active set once the
corresponding absolute correlation reaches the value it would currently
be at had it remained active. Also, while the variable is out of the
active set, it is ignored in the calculation of the maximum absolute
correlation in step 2 of the FLASH algorithm.
\end{definition}

In LARS it is easy to keep track of what the absolute correlation
value would be if the removed variable remained active: it is just
the value of the maximum absolute correlation. In FLASH this value
is also easy to keep track of, because all pairwise ratios among the
active absolute correlations stay fixed throughout the algorithm.

%s9 ###
\section{Path algorithm for the GLM FLASH}

By analogy with LARS and GLasso, the GLM FLASH algorithm progresses in
piecewise linear steps.
Our algorithm is a modification of the one in \citet{park1}.
Throughout the algorithm we write $\blam$ for the vector of absolute
correlations between the predictors and the current residuals,
$|X^T(Y-\hat\mu^l)|$.
We start with $\hat\bbeta=\bolds{0}$, $\hat\mu=\bar{Y}\bolds{1}$, and
the active set, $\A$, consisting of $j^*=\arg\max\lambda_j$.
We decrease the value of $\|\blam_{\A}\|_{\infty}$ from
$|X_{j^*}^T(Y-\bar{Y}\bolds{1})|$ to zero along a data dependent grid.
At each grid point we take the following four steps. The details of
steps 1 and 2 are discussed in \citet{park1}:

\begin{enumerate}

\item\textit{Predictor.} Linearly approximate the solution to (\ref
{wtd.glasso}); call it $\tilde\beta$.

\item\textit{Corrector.} Use $\tilde\beta$ as the warm start to produce
$\hat\beta$, the exact solution to
%
%e6 ###
\begin{equation}
\label{wtd.glasso}
\min_{\bbeta: (\bbeta_{\A^c}) = \bolds{0} } \biggl( -l(\bbeta)+\sum_{j\in
\A}\lambda_j|\beta_j| \biggr).
\end{equation}

\item If $\|\blam_{\A^c}\|_{\infty}\ge\|\blam_{\A}\|_{\infty}$
or $\min_{\A}|\hat\beta_j|=0$, set $\blam_{\A}\leftarrow
(1-\delta)\blam_{\A}$ and
repeat steps 1--2.

\item Let $\A_z$ contain the indices of the zero coefficients in $\A
$. If $\|\blam_{\A^c}\|_{\infty}\ge\|\blam_{\A}\|_{\infty}$,
augment $\A$ with $j^*=\arg\max_{\A^c}\lambda_j$.
Set $\A\leftarrow\A\setminus\A_z$.

\item Set $\blam_{\A}\leftarrow(1-\epsilon)\blam_{\A}$ for some
small $\epsilon$.

\end{enumerate}

Note that setting $\delta=0$ recovers the GLasso algorithm in Park and
Hastie, while setting $\delta=1$ results in the path for GForward.
\end{appendix}

\section*{Acknowledgments}
We would like to thank the Editor, Associate Editor and referees for many
helpful suggestions that improved the paper.

\begin{supplement}[id=suppA]
\stitle{Forward-LASSO with adaptive shrinkage}
\slink[doi]{10.1214/10-AOAS375SUPP} %[doi,text={...}] - jei reikia suskaldyti doi
\slink[url]{http://lib.stat.cmu.edu/aoas/375/supplement.pdf}
\sdatatype{.pdf}
\sdescription{This material contains the proofs of Claim \ref{counter.ex} and Theorem~\ref{vstheorem3}.}
\end{supplement}

\printaddresses

\end{document}